\documentclass{elsart}
\usepackage{graphicx,color}
\usepackage{amsmath}
\usepackage{epsfig}

\begin{document}
\begin{frontmatter}
\title{Summary of Theoretical Physics for ICPAQGP 2010} 

\author{L. McLerran}
\address{Physics Dept. and Riken Brookhaven Research Center 
Bdg. 510A, Brookhaven National Laboratory Upton, NY-11973, USA}
\
\begin{abstract}
This talk includes a discussion of recent theory developments in the areas related to ultra-relativistic
heavy ion collisions.   It includes the topics of the Quark Gluon Plasma, Color Glass Condensate, Glasma and Quarkyonic Matter.\footnote{Presented at the International Conference on the Physics and Astrophysics of the Quark Gluon Plasma, Goa, India, December 5-10, 2011}
\end{abstract}
\end{frontmatter}
  
\section{Introduction}

This lecture will  summarize developments presented at this meeting on topics related to the properties of high energy density strongly interacting matter. This includes the theory of matter in thermal equilibrium such as the Quark Gluon Plasma (QGP),   Quarkyonic Matter and Color Superconducting Matter.  It will also include other forms of matter associated with ultra-relativistic nuclear collisions such a as the Color Glass Condensate  (CGC) and the Glasma.   I have also been asked by the organizers to cover new developments related to the Color Glass Condensate and the Glasma not presented in this conference.

There have been many recent developments in both theory and experiments for the Color Glass Condensate and the Glasma.  These include effects similar to ``jet quenching" that appear for the fragmentation of the deuteron in dA collisions, and the ridge as seen in two particle correlations for both AA and pp collisions.  There is also recent work in the description of pp collisions at LHC energies that suggests that effects related to the CGC may provide quantitative description of the
multiplicity and transverse momentum distributions. 

The Glasma evolves in time from the CGC to the QGP.  I will discuss some very speculative work on how thermalization might occur due to the high degree of coherence of the Glasma fields.

The theory  of the phase diagram of matter at finite temperature and potential has the well known Hadronic Matter, Quark Gluon Plasma and Color Superconducting Phase.  Recent work suggests there is another phase, Quarkyonic Matter.  This Quarkyonic Matter may form a Happy Island in the phase diagram that is surrounded by lines of phase transitions.  There is a triple point where the Hadronic Matter, QGP and Quarkyonic Matter meet, and a possible critical point near the triple point.

Many new results were presented on the lattice gauge theory properties of high energy density matter.  Much of this is related to searching for the critical point referred to in the previous paragraph.

Jet computations are being tested by recent results from the LHC.  In particular recent jet quenching results from LHC heavy ion collisions are consistent with previous descriptions of the phenomenon based on perturbative QCD.  Electromagnetic properties, as measured at RHIC, present challenges for theorists.

I elaborate on the issues outlined above in this talk.  I will refer to the presentations given at this meeting by the name of the presenter.

I apologize for the sparse references that space permits for this summary, and that many very interesting topics are not covered also due to this limitation.  In many cases, original references need to be found in the contributions of plenary speakers of the meeting to whom I refer.

\section{Color Glass Condensate}

The Color Glass Condensate is the matter that is important for high energy collisions involving strongly interacting particles.  It has a very high energy density, $\epsilon >> \Lambda_{QCD}^4$ and is a highly coherent ensemble of states of gluons that form the part of the wavefunction of a hadron that controls typical high energy hadronic collisions.  It can be probed in high energy lepton-hadron collisions  or in high energy hadron-hadron collisions.  The CGC is parameterized by a scale, the saturation momentum,
that is related to the density of gluons as
\begin{equation}
  {{dN} \over {dyd^2r_T}} \sim {1 \over \alpha_S} Q_{sat}^2
\end{equation}
In most phenomenological analysis of hadronic collisions, the density of gluons is taken to be proportional to the final state density of hadrons.  The phase space occupation density of gluons
for $p_T \le Q_{sat}$ is very large
\begin{equation}
  {{dN} \over {dyd^2r_T d^2p_T}} \sim {1\over \alpha_S}
  \end{equation}
This means that the gluons are highly coherent, and that they can be described by a classical field.

The CGC provides a very good description of deep inelastic scattering for $x \le 10^{-2}$ for $ep$ collisions as seen at the HERA accelerator.  It might be tested in $eA$ collisions at an electron ion collider.  One of its simple predictions is that the cross section for deep inelastic scattering scales
$\sigma_{\gamma^* p} = F(Q^2/Q_{sat}^2)$\cite{Stasto:2000er}.  This   is well tested in the HERA data and provides an excellent description when 
\begin{equation}
Q_{sat}^2 = Q_o^2 (x_o/x)^\lambda
\label{qsat}
\end{equation}
with $\lambda \sim 0.2 - 0.3$

It appears that the data on pp scaling satisfies a similar scaling relation
\begin{equation}
  {{dN} \over {dyd^2p_T}} = F \left( {p_T \over {Q_{sat}(p_T/\sqrt{s})}} \right)
\end{equation}
\begin{figure}[t]
 \center{\vskip 0in \hskip 0in\includegraphics[width=14cm]{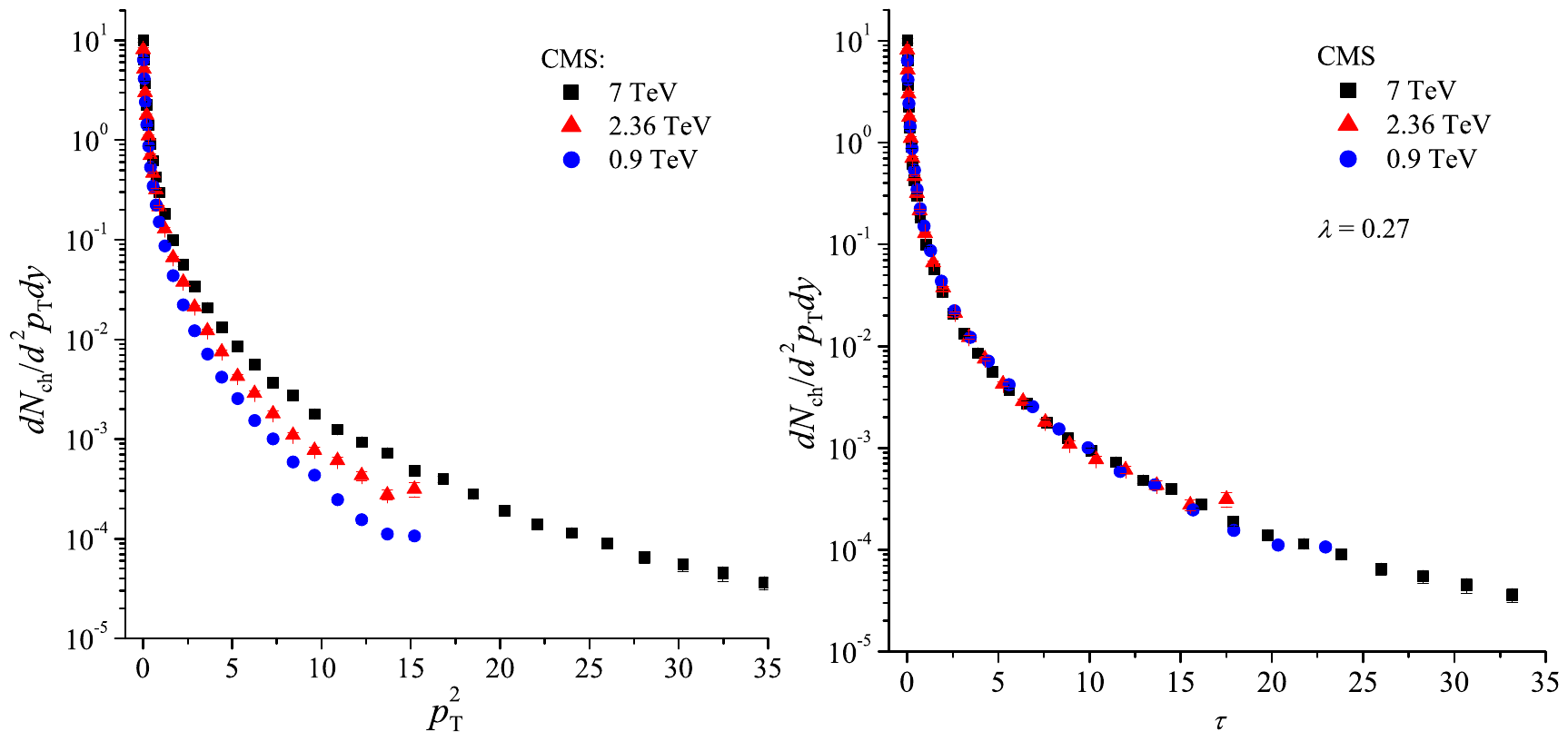} \vskip 0in }
 \caption{\label{gscal_pp}  On the left,  transverse momentum particle distributions
 as measured at LHC with the CMS detector, and on the right plot, compared with a scaling fit }
 \end{figure}
Taking the CMS data from LHC\cite{Khachatryan:2010xs}-\cite{Khachatryan:2010us},
 one can check scaling\cite{McLerran:2010ex}-\cite{McLerran:2010wm}.  For the choice 
 of $\lambda = 0.27$ in Eqn. \ref{qsat}, there is excellent scaling.  Such a scaling extends down to SPS and RHIC energies.  One can also take a first principle computation based on saturation models that have well described that data, and as was shown by Tribedy at this meeting, obtain a good description of the LHC data\cite{Tribedy:2011yn},
\begin{figure}[t]
 \center{\vskip 0in \hskip 0in\includegraphics[width=14cm]{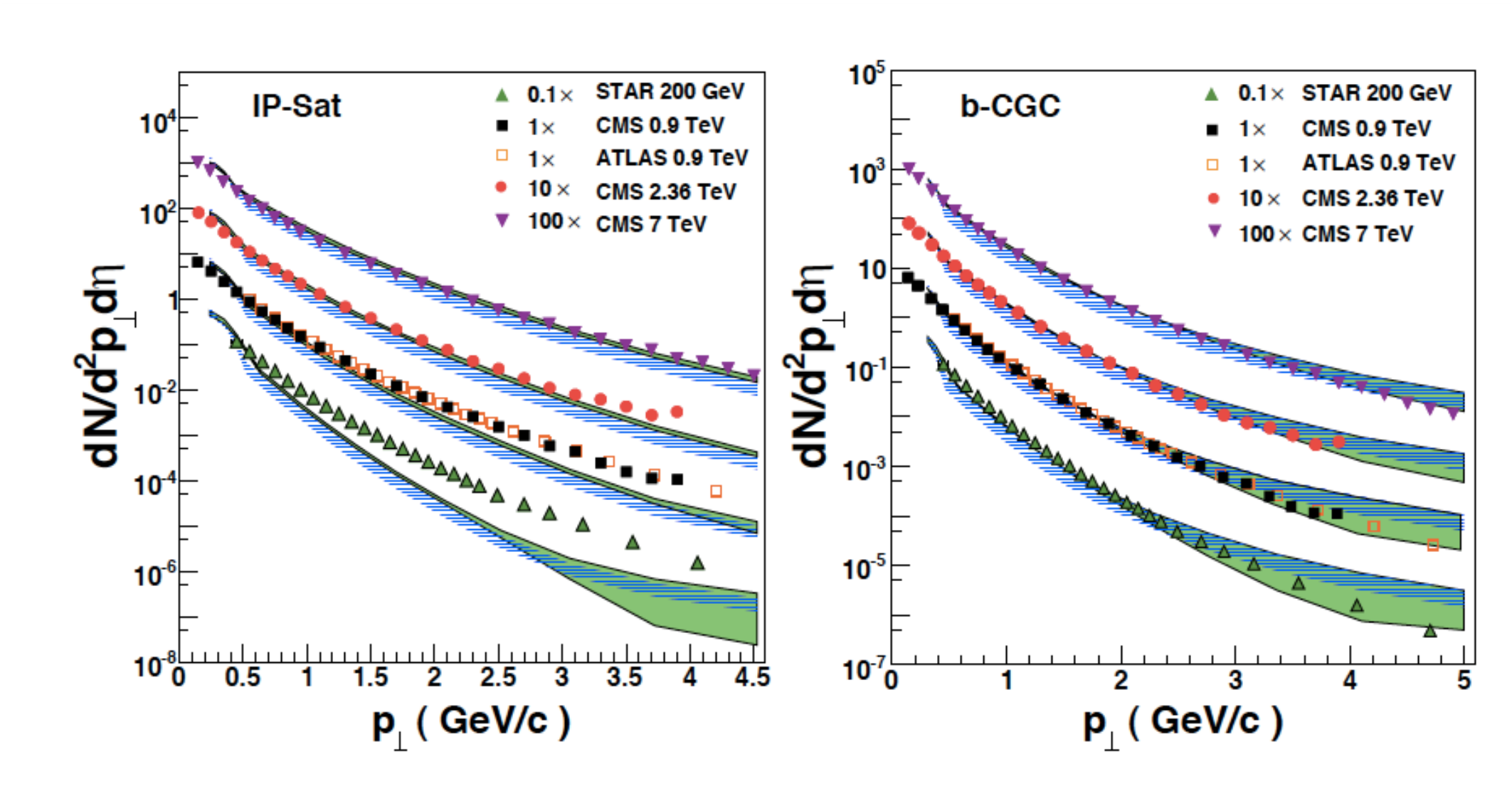} \vskip 0in }
 \caption{\label{tribedy_pp}  A description of LHC pp data\cite{Abe:1988yu}, \cite{Khachatryan:2010us},\cite{Aamodt:2010pp},
 using saturation models.  See  \cite{Tribedy:2011yn} for references to the saturation models.}
 \end{figure}
 
 As a further test of the CGC description, one predicts that the average transverse momentum squared will scale as the multiplicity in multiplicity fluctuations.\cite{McLerran:2010wm}  This is observed in the Atlas experiment\cite{Aad:2010rd}.
 
\begin{figure}[t]
 \center{\vskip 0in \hskip 0in\includegraphics[width=14cm]{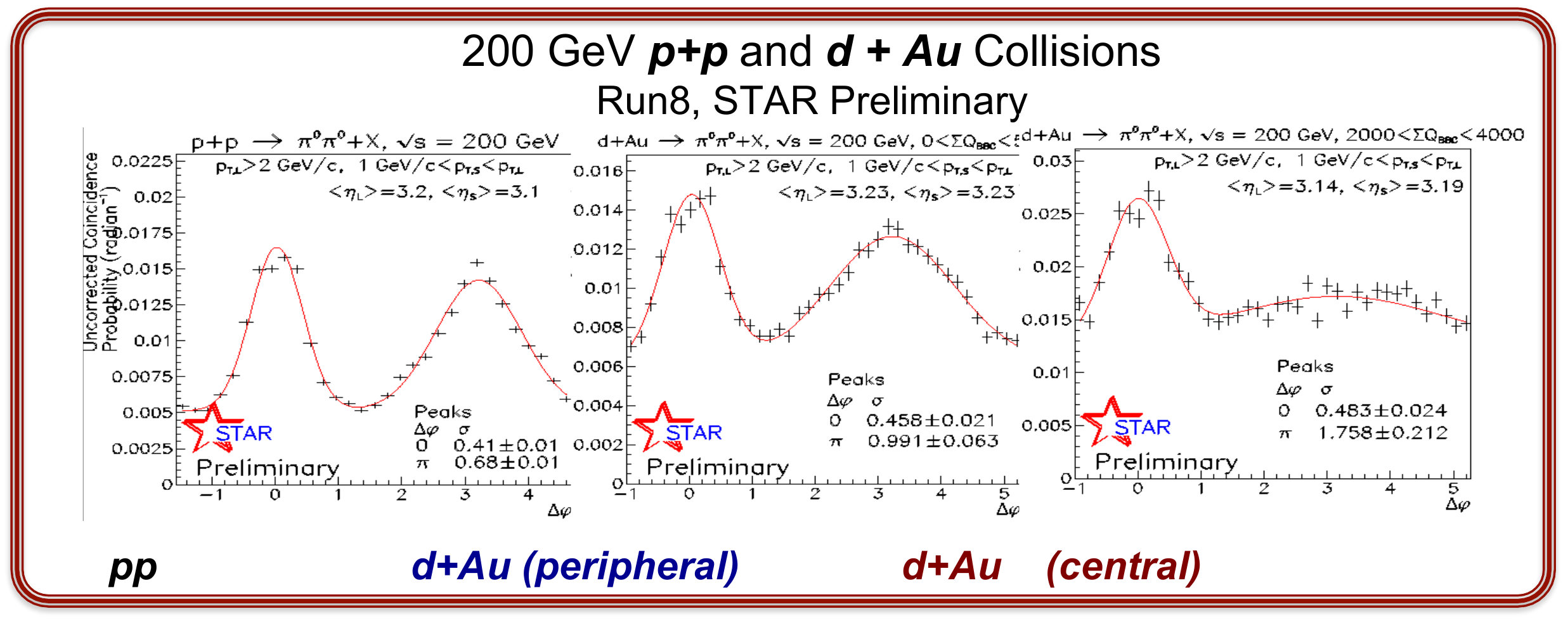} \vskip 0in }
 \caption{\label{dA}  STAR data on two particle correlation in dAu colliisions}
 \end{figure}

 Collisions of deuterons from nuclei provide a means to probe the CGC.  The fragmentation region of the deuteron probes the small x part of a nuclear wavefunction.  If the collision produces a high transverse momentum particle by an elementary process, there should be a recoil particle in the backward direction.  On the other hand, a CGC can absorb the recoil momentum of such a backwards going particle, and the recoil peak as seen at fixed transverse momentum will be diminished.  Such a  reduction was seen in the star experiment \cite{Braidot:2010zh}, as shown in Fig. \ref{dA}.  
This data is well described within CGC models\cite{Albacete:2010pg}-\cite{Tuchin:2009nf}.
No such backward suppression was seen at central rapidity in dAu collisions, as was discussed by Prindle at this meeting.

\begin{figure}[t]
 \center{\vskip 0in \hskip 0in\includegraphics[width=14cm]{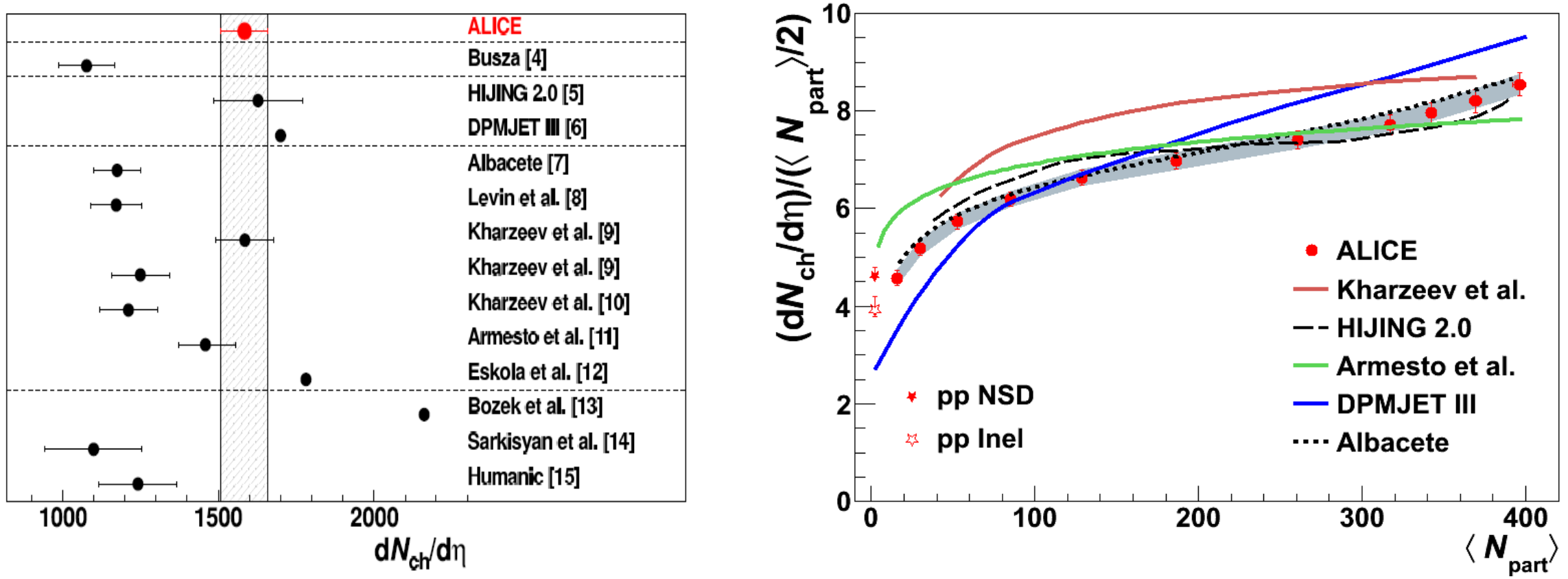} \vskip 0in }
 \caption{\label{mult_pbpb}  The total multiplicity measured in Alice and the dependence of the multiplicity upon energy.}
 \end{figure}

Alice has presented data on the total multiplicity in Pb-Pb collisions at 2.76 TeV \cite{Aamodt:2010pb}.  This presents a test of saturation models.  In Fig. \ref{mult_pbpb} both the measured multiplicity and the dependence of the multiplicity on centrality are shown compared to a variety of models.  The saturation based models include those labeled as  Hijing, Albacete, Levin, and Kharzeev (please see the original paper for references to these models).  They tend to be a bit on the low side.  There is undoubtedly an effect of theorists being unable to properly estimate their systematic errors.  (Some theorists believe their systematic errors are zero until confronted with reality.)  For the dependence of the multiplicity on centrality, up to an overall normalization, the general dependence is reasonable.  (DPMJET III is not a saturation based model.)  It is no doubt reasonably simple to correct the problems in the saturation models used to predict the total multiplicity,  and one will learn from the corrections that need be done,  but it is disappointing that the predictions were not better.  

\section{Glasma}

\begin{figure}[t]
 \center{\vskip 0in \hskip 0in\includegraphics[width=14cm]{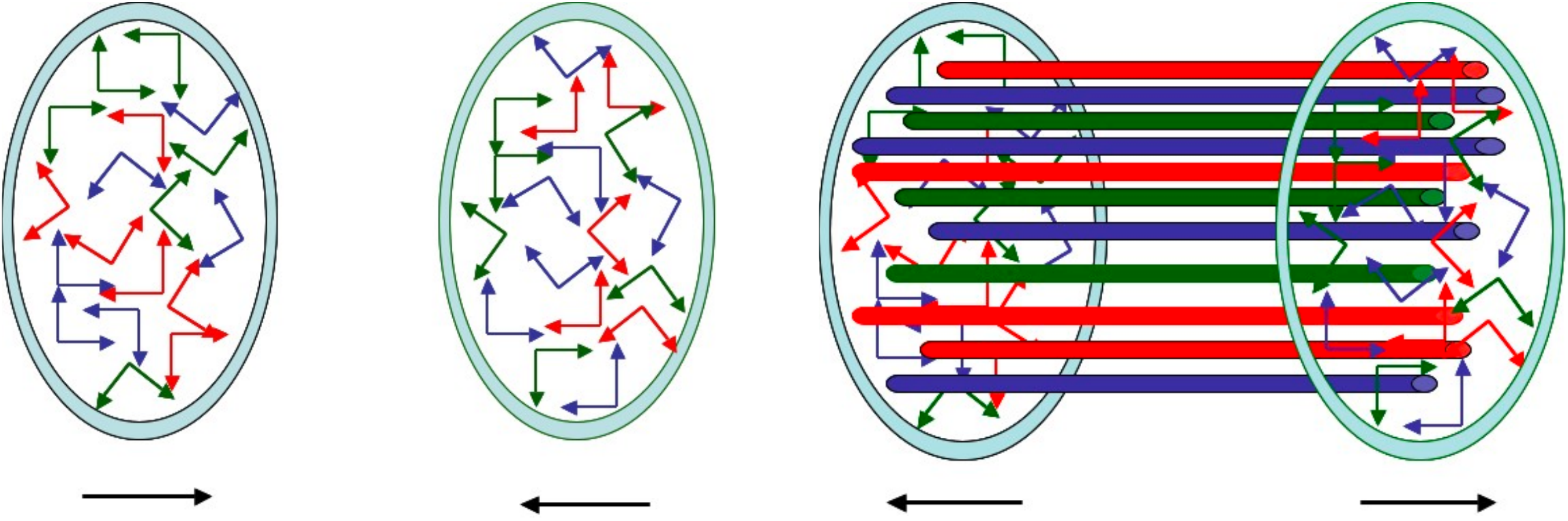} \vskip 0in }
 \caption{\label{glasma}  The CGC before the collision of two nuclei, and the Glasma formed after their collision.}
 \end{figure}

The distribution of colored fields of CGC, corresponding to a high energy nucleus, are color electric and color magnetic fields propagating near light speed in a thin sheet.  These color electric and color magnetic fields are plane polarized perpendicular to the direction of motion of the nucleus.  This is shown in Fig. \ref{glasma}.  After these sheets pass through one another,  longitudinal color electric and color magnetic fields are formed.  This is because when the nuclei pass through one another, they are dusted with color electric and color magnetic charge.  The longitudinal fields arise because of Guass's law and the geometry of the collision.  Note that the color electric and magnetic field have non-zero $\vec{E} \cdot \vec{B}$ corresponding to a topological charge density.  The fields are initially very strong and highly coherent.  The typical transverse size of a line of color electric or magnetic flux is of order $r \sim 1/Q_{sat}$, and the typical electric or magnetic field has strength $Q_{sat}^2/g_{S}$. As time evolves, these Glasma fields decay into quarks and gluons, and they eventually form a thermalized QGP.
 
Thermalization of the Glasma has not yet been solved.  It is believed that very early in the collision,
the fields rapidly become isotropic.  This is plausible because the fields begin in a small corner of phase space,and typically such fields will explore other regions of phase space by developing turbulence or through parametrically amplifying modes.  Explicit solutions of scalar field theory exhibit such behavour\cite{Dusling:2010rm}.  We can understand this in classical mechanics: If we have a collection of large number of ball with momenta all in one direction, it takes a few collisions per ball to make the momentum distribution isotropic.

Once the distribution of modes fill phase space and attain approximate isotropy, it is a challenge to maintain itself in with strong self interactions until thermalization.  If this does not occur isotropy is lost, and the momentum distributions of modes will adjust to expansion as would a thermalized system.   In Sinha's talk, thermalization is argued to be  a consequence of strong coupling, yet as one goes to very high energies, the coupling should become weaker, and such argumentation would suggest there is some intermediate time where isotropy and thermalization is lost..  Early data on flow, presented here by Schukraft, suggest that the matter at LHC energies is well thermalized,
which might run against this hypothesis.  Another possibility is that the high degree of coherence
of the classical fields enhances interaction strengths and is responsible for thermalization.  After all, the system starts in highly coherent strong fields, and gluons are evaporated from such matter.
In any case, there is no agreement yet about how thremalization should work nor an explicit scenario that demonstrates it.

\begin{figure}[t]
 \center{\vskip 0in \hskip 0in\includegraphics[width=14cm]{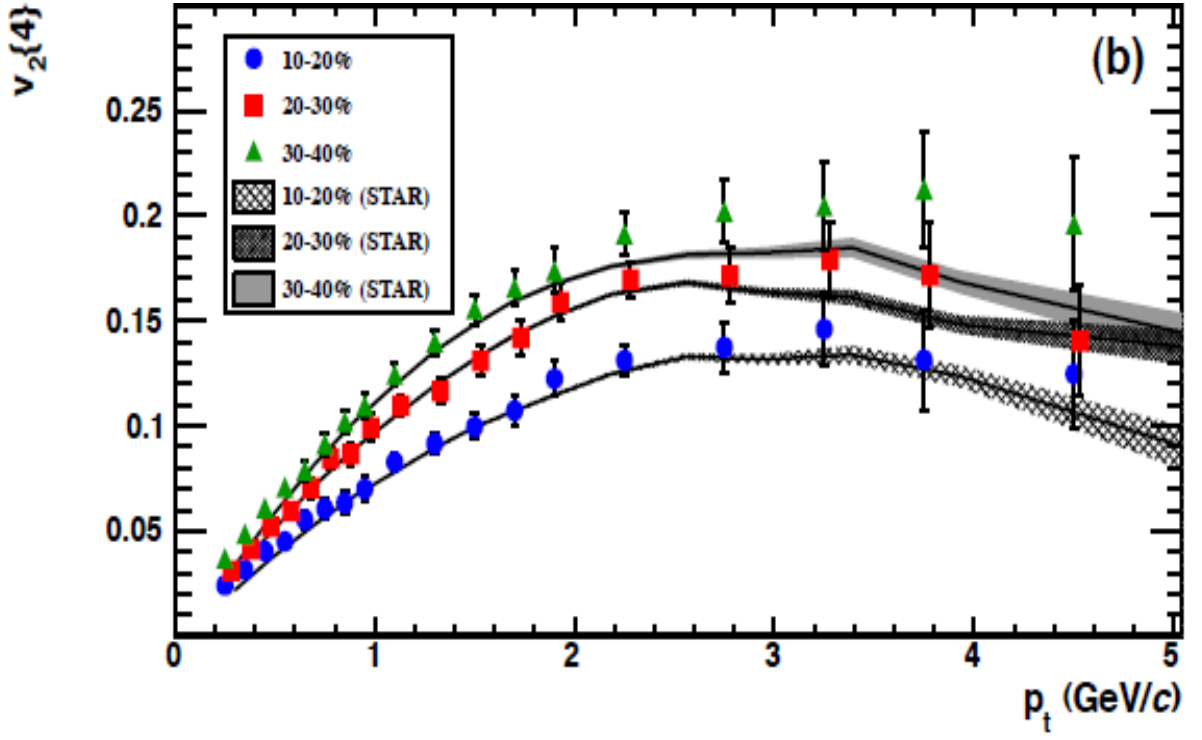} \vskip 0in }
 \caption{\label{flow}  Flow as measured by $v_2$ at RHIC and at LHC.}
 \end{figure}

Certainly the strong flow patterns of seen in the RHIC and LHC experiments reflects an underlying
strongly self-interacting QGP, as shown in Fig. \ref{flow}.  The issue of whether or not this is due to a truly strong coupling or whether it is due to amplification of intrinsic weak coupling dynamics by coherence effects is in my opinion not resolved.  It is argued in the talk by Sinha that intrinsic strong interactions may be modeled using techniques derived from strongly interacting conformal supersymmetric field theory using the  AdSCFT
correspondence.  This correspondence in its simplest form argues there is a lower bound for the viscosity to entropy ratio\cite{Policastro:2001yc}.
\begin{equation}
  {\eta \over S} \ge {1 \over {4\pi}}
\end{equation}
Hydrodynamic simulations of heavy ion collisions suggest that the preferred value is within a factor of two or so of this lower bound.  It might also be the case, that the flow patterns are due to an anomalously small viscosity due to the strong color fields of the Glasma as argued by Bass\cite{Asakawa:2010xf}.
It is difficult to resolve this issue theoretically as the AdSCFT based computations cannot be reliably performed in a realistic theory of strong interactions, and the computations involving strong colored fields have yet to make a convincing case that such fields thermalize into a QGP, and
would also need an explicit computation of such thermalization.
 
 The flux tubes predicted by the Glasma will show up in two particle correlations as described in the talks by Prindle, Schukraft, Shukla and B, Srivastava,  The STAR experiment has observed long range correlation in rapidity fluctuations\cite{:2009dqa}, correlations that are near the maximal allowed value based on very general considerations\cite{Lappi:2009vb}.  Such correlations might arise from Glasma flux tubes.
 
 Perhaps one is directly measuring the long range correlations associated with flux tubes in the ridge measurements first seen in STAR at RHIC\cite{Adams:2004pa}-\cite{Alver:2009id}.  The ridge is a long range rapidity correlation and correlation in azimuthal angle.  Long range rapidity correlations must be established very early in the collision.\cite{Dumitru:2008wn}  The azimuthal angular correlation might be generated either at the emission from a flux tube or later by hydrodynamic or opacity effects.  There is general consensus that the ridge as seen in nucleus nucleus collisions is due to long range tube like structures, and that the inclusive ridge as seen
 in heavy ion collisions is a hydrodynamic effect associated with transverse spatial inhomogeneities associated with the ends of the tubes.  This is quantified in a flow coefficient $v_3$\cite{Alver:2010gr}.  There is not consensus about the origin of  angular structure associated with the ridge triggered  by a high momentum particle.  
 
 The ridge is very clearly seen in the inclusive data for the Alice experiment with PbPb collisions
 as was shown in the talk by Schukraft.  It is also seen in high multiplicity pp collisons\cite{Khachatryan:2010gv}.  The existence of the ridge in pp collisions forces one to conclude that the 
 tubular structures generating the ridge have a transverse size scale that is sub-nucleonic, presumably from sources of quarks and gluons, as predicted in the Glasma description.
 \begin{figure}[t]
 \center{\vskip 0in \hskip 0in\includegraphics[width=14cm]{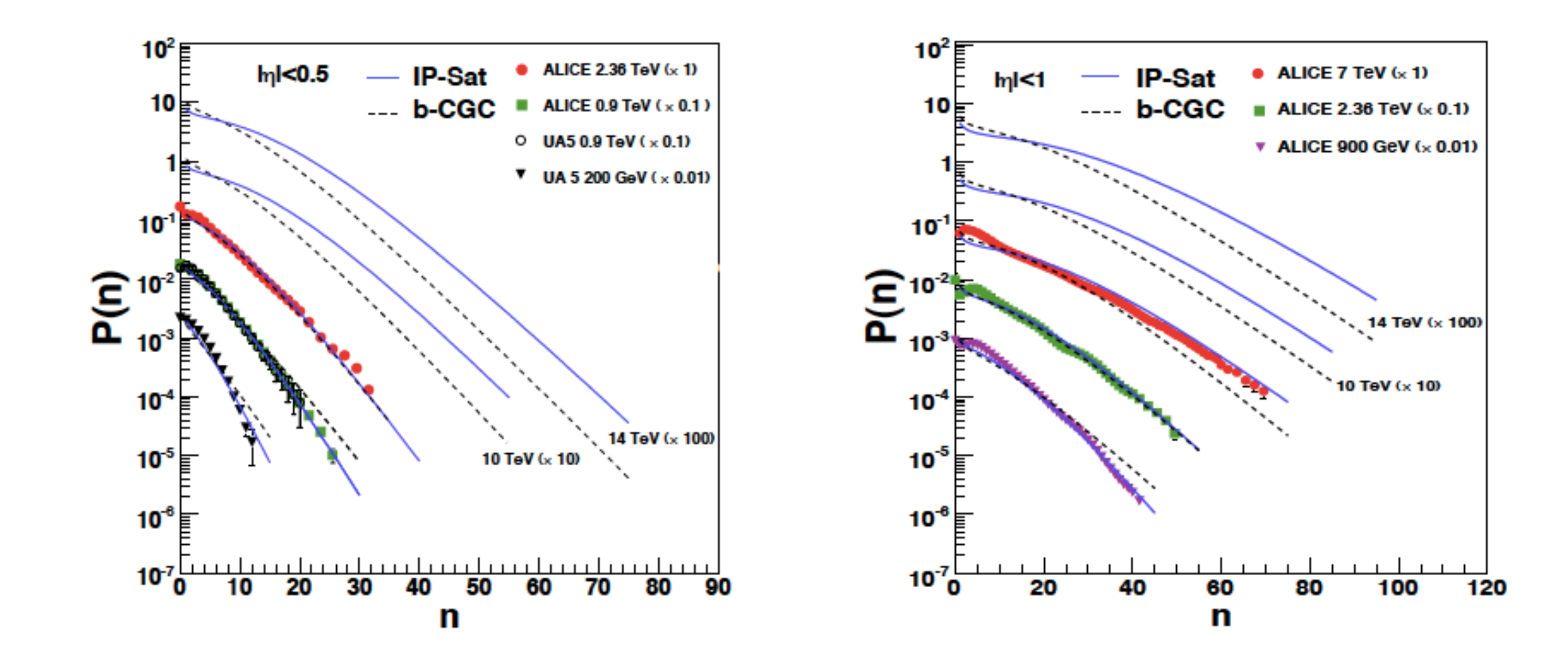} \vskip 0in }
 \caption{\label{glitter}  The Glasma description of the distribution of multiplicities seen in pp collisions by the Alice experiment}
 \end{figure}
 
 When Glasma flux tubes decay, they produce a negative binomial distribution\cite{Gelis:2009wh}.
 The sum of emissions from a negative binomial distribution generates a negative binomial distribution.
 In the talk by Tribedy\cite{Tribedy:2011yn}, it was shown that the decays of such tubes produces a distribution of 
 multiplicities that well describes the LHC data, as shown in Fig. \ref{glitter}
 
 \section{The Phase Diagram of QCD}
  \begin{figure}[t]
 \center{\vskip 0in \hskip 0in\includegraphics[width=14cm]{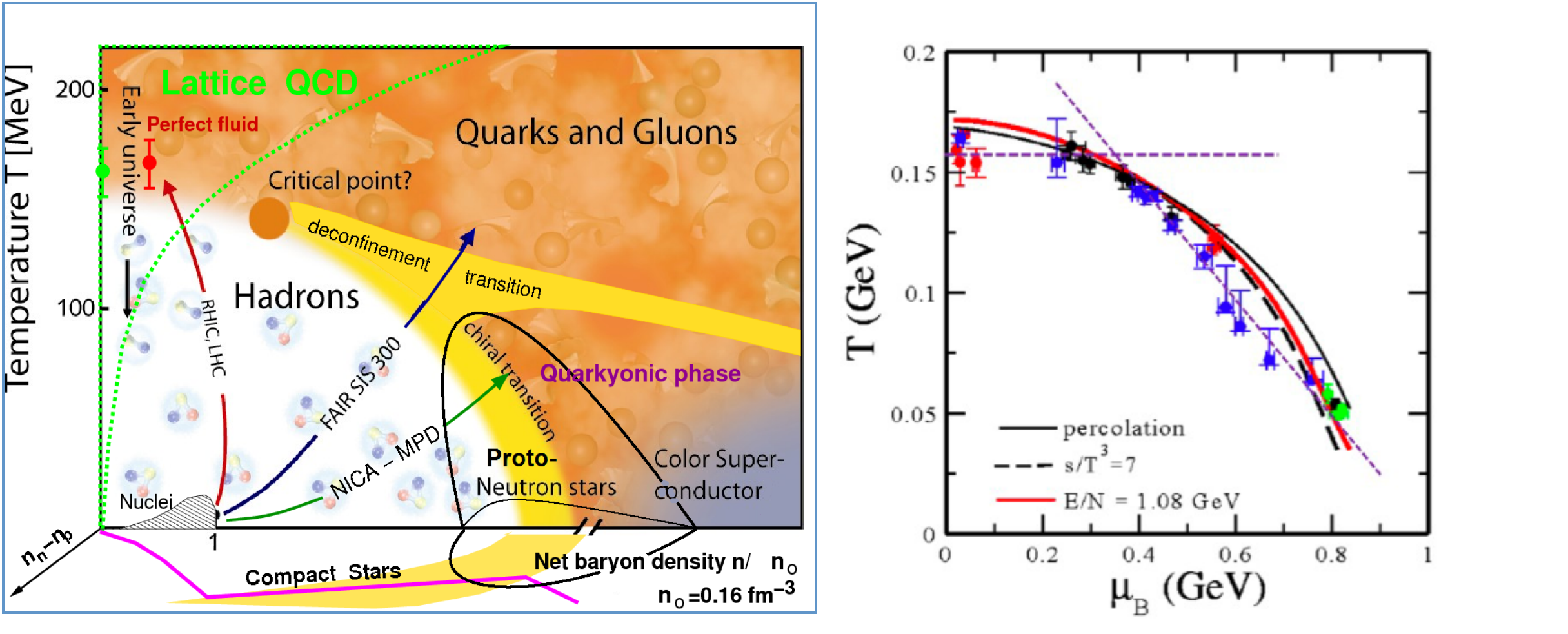} \vskip 0in }
 \caption{\label{tmudiagram}  On the left is a hypothetical phase diagram of QCD.  On the right is
 a the decoupling curve for matter produced in heavy ion collisions as inferred from various experimental measurements }
 \end{figure}
 
 In Stocker's presentation, a  hypothetical phase diagram of QCD is shown, Fig. \ref{tmudiagram}.  Included are the familiar Quark Gluon Plasma, Hadronic Matter and Color Superconducting phases.  Also included is another phase, Quarkyonic Matter.
 
 Quarkyonic Matter is inferred from large number of colors, $N_c \rightarrow \infty$, arguments\cite{McLerran:2007qj}.
 In this limit, the baryon mass is $M_B \sim N_c \Lambda_{QCD}$, so that in the hadronic phase,
 $T \le \Lambda_{QCD}$ there are no baryons.  Baryons may appear at high temperature when chiral symmetry is restored $M_B \rightarrow 0$., or because a baryon chemical potential larger that the nucleon mass is introduced that forces baryon number in the system.
 
 As baryon number is increased, its effects on the confining potential are of order $1/N_c$.  Therefore there can be a phase of low temperature high baryon number density matter where the energy density is parametrically large, $\epsilon >> \Lambda_{QCD}^4$, but is still confined.  Deep inside the Fermi sea interactions are at scales large compared to the confining scale.  But Fermi surface excitations and thermal excitations such as anti-baryons and mesons are confined.  Such matter has properties dramatically different from either that of Hadronic matter of of the Quark Gluon Plasma
 
 The deconfinement temperature is in leading order not affected by the presence of baryons. The transition to Quarkyonic Matter occurs at the chemical potential where baryons first appear and temperatures below that of deconfinement.  
 
 Chiral symmetry is probably broken in a translationally non-invariant chiral spiral\cite{Kojo:2009ha}.  The spiral structure arises because of Pauli blocking.  This means that Quarkyonic Matter very probably breaks translational invariance, and if so it is surrounded by a line of phase transitions in the $\mu -T$ plane.  I call this region Happy Island.
 
\begin{figure}[t]
 \center{\vskip 0in \hskip 0in\includegraphics[width=14cm]{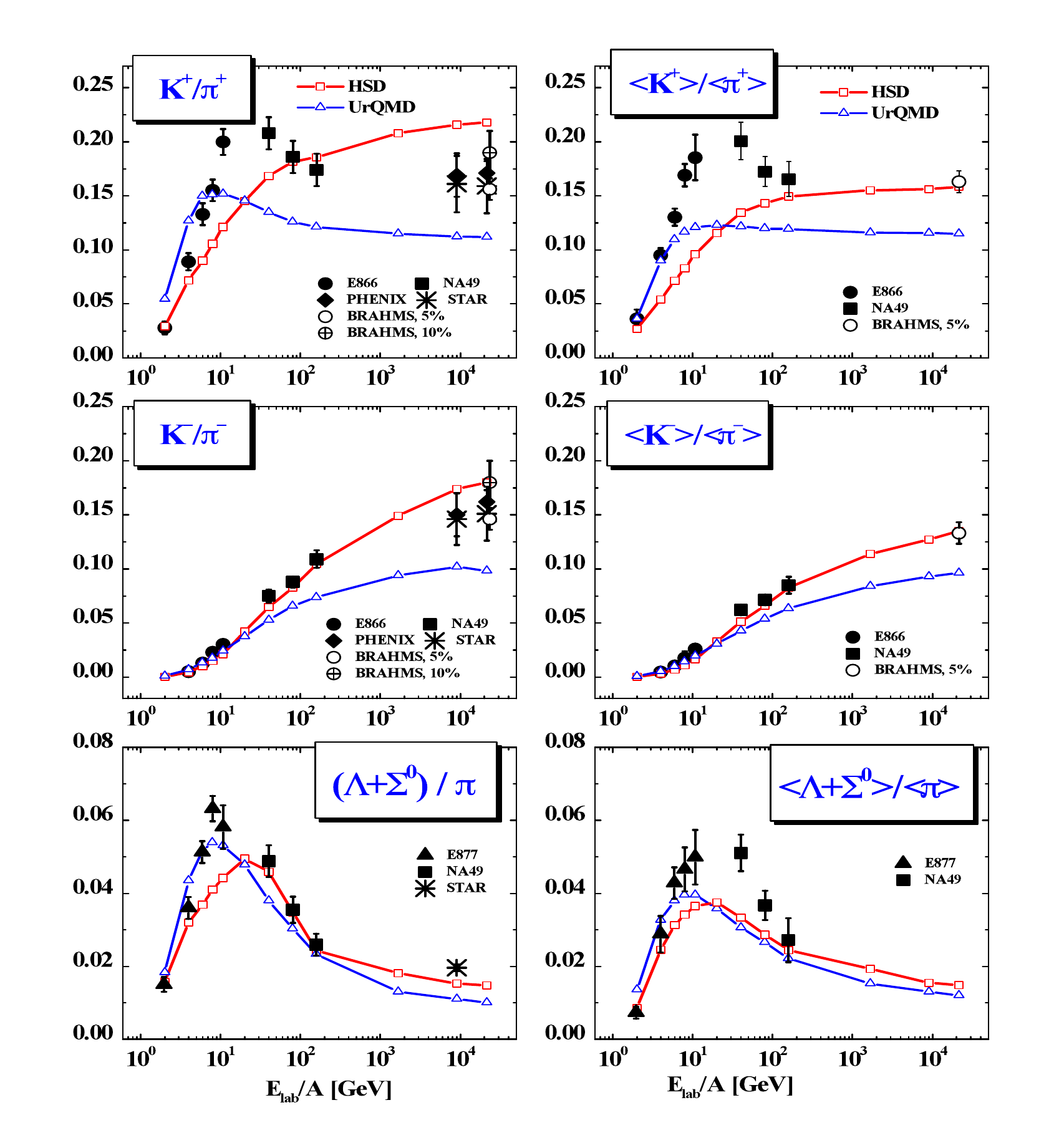} \vskip 0in }
 \caption{\label{marek}  Ratios of particle abundances as a function of energy.}
 \end{figure}

 The phase deconfinement transition and the transition to Quarkyonic Matter may be inferred from
 the decoupling of the rapidly expanding matter produced in heavy ion collisions.  This is because both the transition of the Quark Gluon Plasma and Quarkyonic matter to a Hadron Gas involve very large changes in the number of degrees of freedom.  As these degrees of freedom are diluted by expanding the  energy density, the system remains at a fixed temperature and chemical potential.
 The results of such considerations are shown in Fig. \ref{tmudiagram}, and the original papers on this subject are referred to in Ref. \cite{Andronic:2009gj}.  Shown on the plot are very simple models of the confinement and Quarkyonic transitions, the dashed lines.  There is a point where
 the three phases coexist, the triple point, and the triple point may explain a number phenomena seen in experiments that probe this region of $\mu-T$.  This is shown in the ratios of particle abundances in Fig. \ref{marek}
 
 A critical end point may also appear in the phase diagram, presumably at somewhat lower baryon chemical potential below the triple point\cite{Rajagopal:1992qz}.  As discussed by Gavai and S. Gupta, lattice gauge theory computations can search for this breakdown of perturbation theory in an attempt to find a first principles determination of the critical end point.
 
 Susceptibilities are measures of the response of the system to changes in chemical potential and temperature, and can be measured near the deconfinement or Quarkyonic transitions.   Near such transition one might expect singular behavour, and as such provide a signal for a phase transition.
 of particular interest is the behavour near a critical end point. They may be computed from first principle from lattice Monte Carlo computations
 or from hadron resonance gas models.  In the talks of Gupta and Redlich, we saw that the computations of these quantities at high temperature, and not so large baryon chemical potential are quite reliable.  Comparisons were made with data and provide remarkable agreement,
 although experimental signature for a critical end point  was not found.
 
 The properties of matter near the Quarkyonic region might be studied in the RHIC low energy run, at FAIR with the CBM detector and at NICA.  NICA provides a dedicated collider facility with plans for a detector that can provide unique information on the central region of heavy ion collisions in the region where one expects effects of a triple point and critical end point.

 \section{Lattice Gauge Theory}
 
 The talks of Gavai, R. Gupta and S. Gupta summarized the status of lattice gauge theory computation.  The results on matter at low baryon density and high temperature are very impressive and there is consensus on numbers, within systematic and statistical uncertainty, of computations using different methods.  Computations allow for a reliable extraction of the equation of state by combing lattice Monte Carlo data with hadron resonance gas models\cite{Huovinen:2009yb}.  These equations of state may be used as input in hydrodynamic simulations of heavy ion collisions.
 
 One of the challenges of lattice gauge theory, as described by Gavai, is to do reliable computations at high baryon number density.  One can extrapolate into the $\mu-T$ plane to values $\mu_B/T \sim 1$  Although there is tantalizing suggestions that one might be seeing the effects of a critical end point, there is not yet consensus in the lattice gauge theory community that numbers may be reliably and unambiguously extracted. 
 
 \section{Jets and High $p_T$ particle Production}
 
  \begin{figure}[t]
 \center{\vskip 0in \hskip 0in\includegraphics[width=14cm]{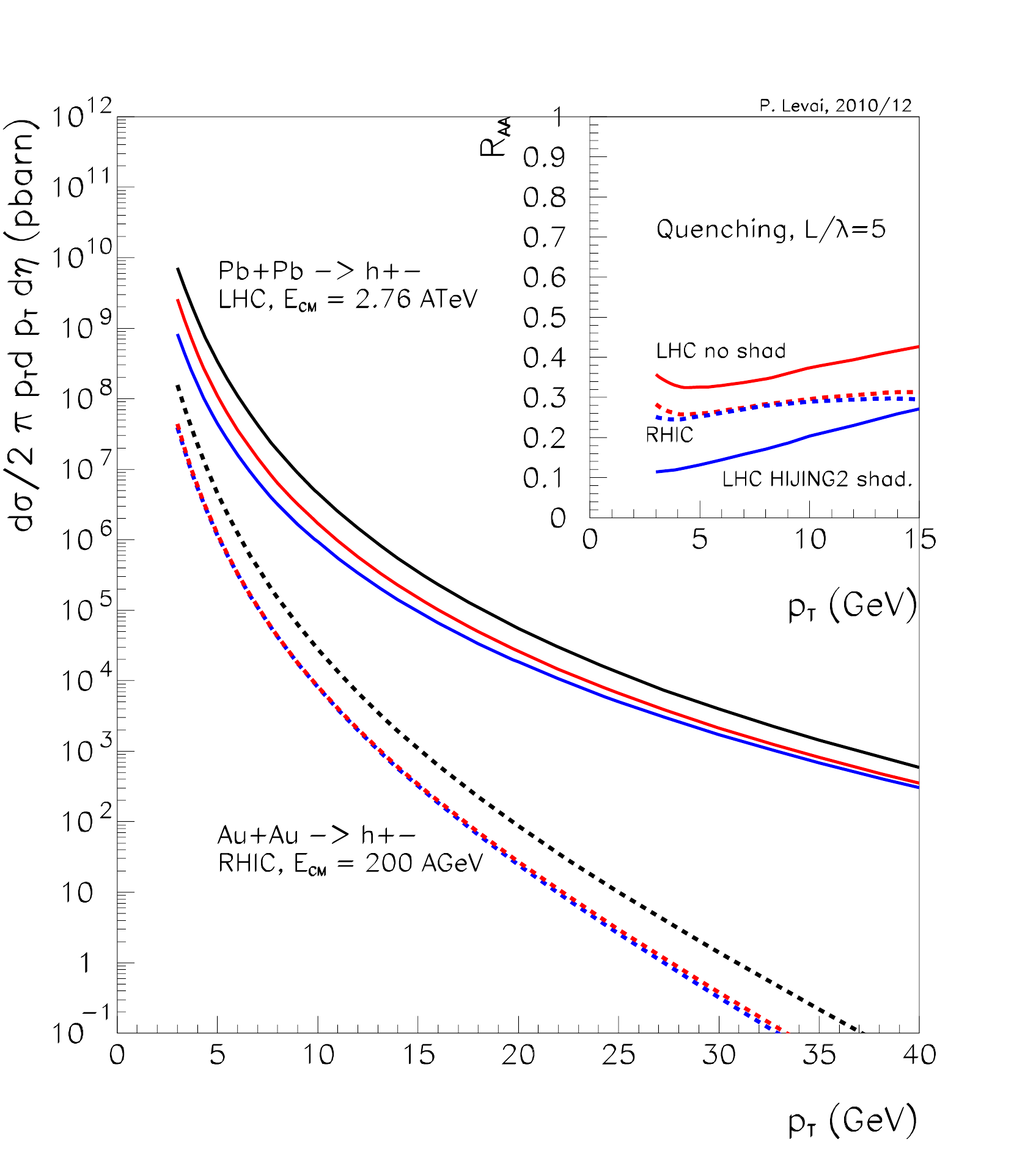} \vskip 0in }
 \caption{\label{levai}  Charged particle distribution at RHIC and at LHC in nucleus nucleus collisions.  The ration $R_{AA}$ is shown in the solid lines for computations at LHC energy that have only jet quenching (upper curve) and jet quenching with shadowing (lower curve).  The computed jet quenching at RHIC is shown in the dotted curves.}
 \end{figure}
Exciting results and their theoretical interpretation concerning hard particle production at LHC were shown in the talks by Levai, Schukraft, Shukla, Giubellino and Montoya.  At first sight, the results seem a little mysterious as the data at LHC show a larger ratio $R_{AA}$ than is the case at RHIC, which might naively suggest less jet quenching.  Levai provides a good explanation for this shown in Fig. \ref{levai}.  The point is that for the same down shift in jet energy at LHC compared to RHIC, one generates a smaller shift in $R_{AA}$, .  This is because the fall off of single particle distribution at LHC is less rapid than at RHIC.  In order
to explain the data at LHC within a conventional jet quenching, one needs much more shadowing than at RHIC (where to a first approximation shadowing was not required).   The bottom line is that
the description advocated for ``jet quenching'' at RHIC energies has (so far) little trouble in describing the LHC  data.
  \begin{figure}[t]
 \center{\vskip 0in \hskip 0in\includegraphics[width=14cm]{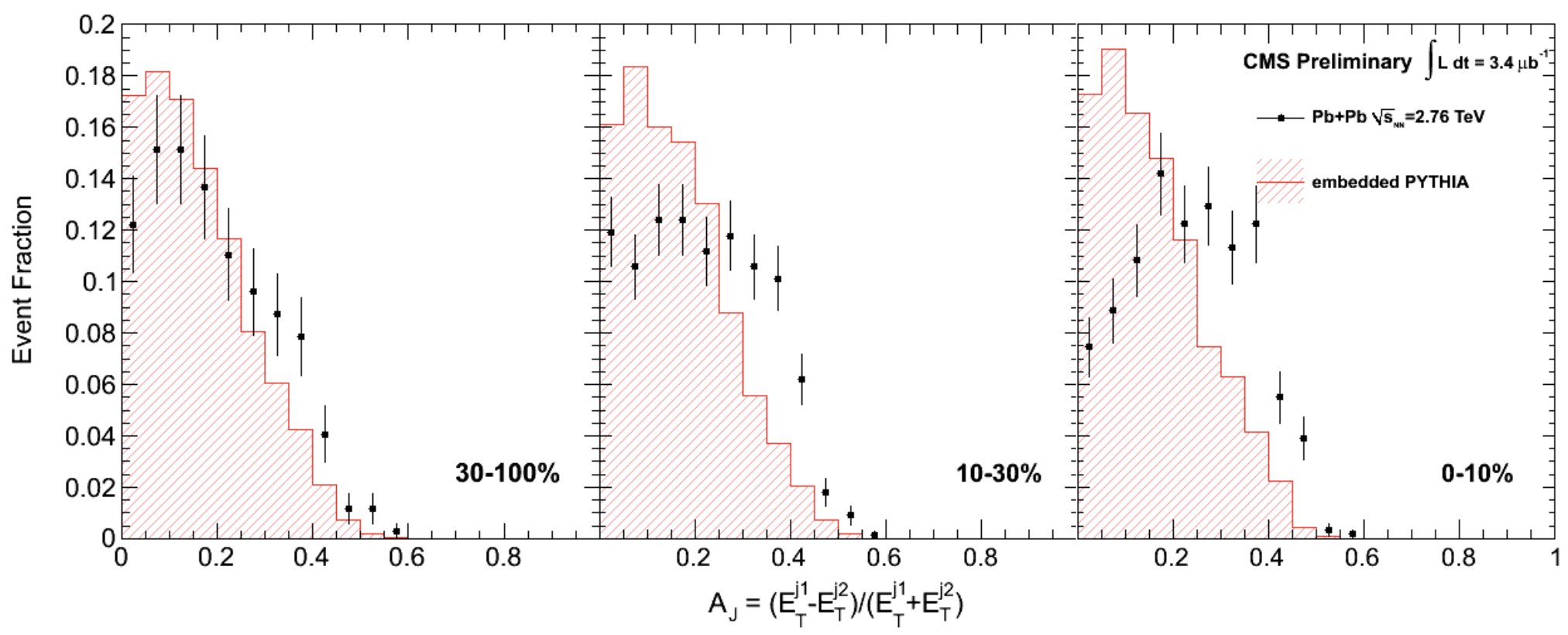} \vskip 0in }
 \caption{\label{cms}  Jet asymmetry energy scaled to total energy of jets in CMS experiment}
 \end{figure}
 
 Atlas and CMS presented data on jet quenching in heavy ion collisions.  In the CMS plot, the asymmetry energy of two jets is scaled to the total energy of the two jets plotted relative to the total energy of two jets in the CMS experiment as a function of centrality  in heavy ion collision.  There is energy loss but it appears to be small relative to the total energy of the jet.  This is qualitatively dissimilar to predictions based on strong coupling models.  It appears that the CMS and Atlas experiments are consistent with a conventional jet energy lost mechanism with\cite{CasalderreySolana:2010eh}
 \begin{equation}
  3~GeV^2/Fm < \hat{q} < 10~GeV^2/Fm
 \end{equation}

\section{Electromagnetic Probes af the Quark Gluon Plasma}

\begin{figure}[t]
 \center{\vskip 0in \hskip 0in\includegraphics[width=14cm]{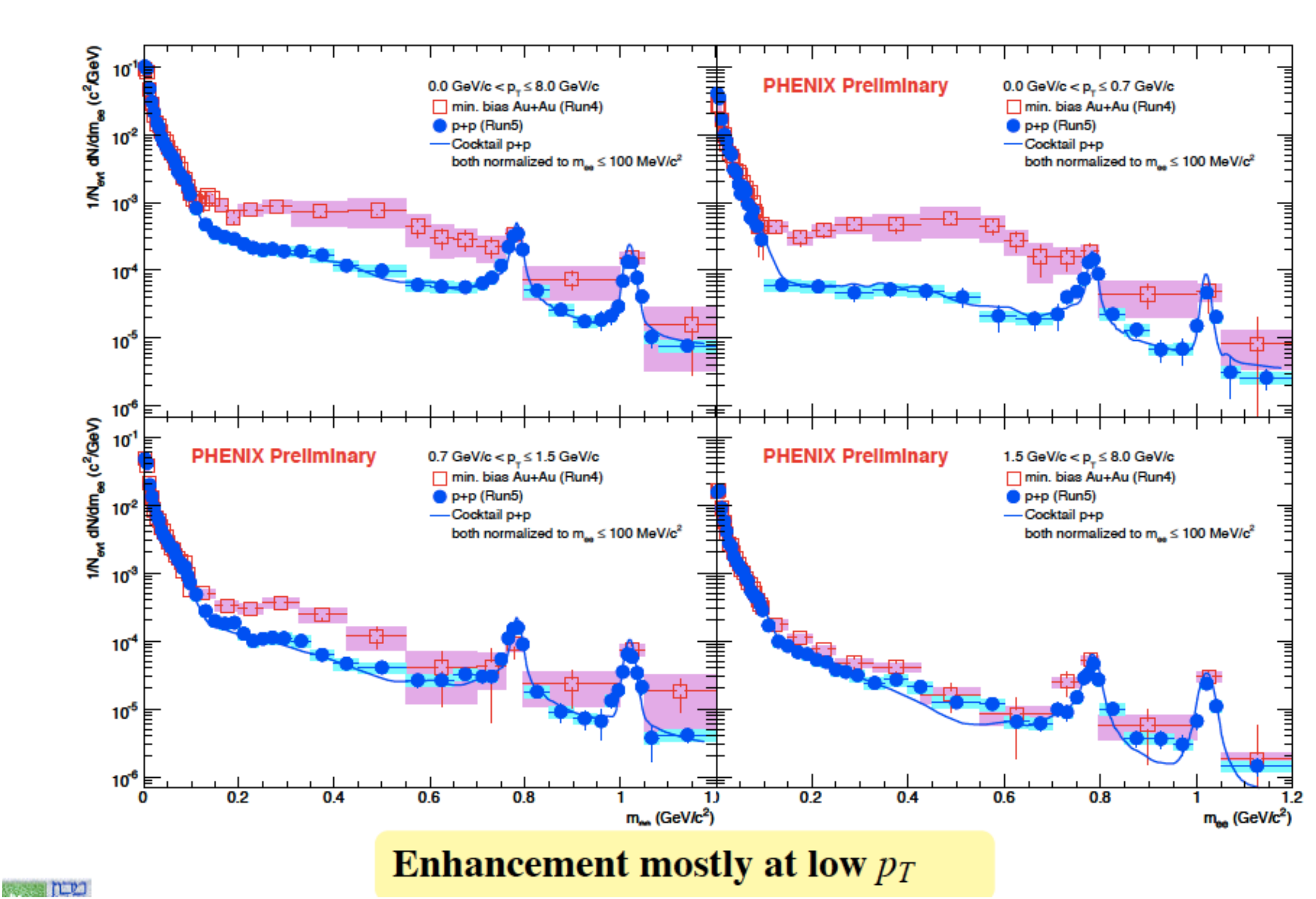} \vskip 0in }
 \caption{\label{phenixee}  Yields of di-lepton pairs as a function of dilpeton mass}
 \end{figure}

Talks were presented by Rapp, Jan-e Alam, Sharma, Stroth, Braun-Munzinger and Noucier concerning electromagnetic probes of the QGP.  Phenix results  on dileptons have an enhancement in the mass range of $200~MeV \le M_{ee} \le 700~MeV$\cite{Adare:2009qk}, as shown in Fig. \ref{phenixee}.
 This kinematic region is roughly where one might expect effects of a hot Quark Gluon Plasma
 would begin to become important.  The effect has so far evaded theoretical description because it is rather large and it peaks in the low transverse momentum region.
 
 There was also much discussion of the low $p_T$ enhancement of photons seen in Phenix\cite{:2008fqa}.  This is for photons with $p_T \le 3-4~GeV$.  This can be successfully computed in
 hydrodynamic simulations.  Such simulation typcially have an initial temperature in the range of $300~MeV$ to $600~MeV$.  The slope of the $p_T$ distribution of such photons corresponds to a
 thermal distribution with a temperature of $221\pm19(stat) \pm19(sys)$ although the direct interpretation of this as a temperature is complicated since the photons are emitted from an expanding thermal distribution with a time varying temperature.  It is not clear from theoretical considerations what non-thermal mechanisms might be important. Nevertheless, the typical energy scale associated with these photons is above that of the de-confinement transition.
 
 There was much discussion of $J/\Psi$ production and heavy quarks.  This area of research remains theoretically challenging.  Particularly difficult to explain is the large energy loss and the robust flow patterns inferred for heavy quarks.  A variety of new measurements at RHIC and LHC involving direct measurements of charm may help to clarify understanding.
 
 \section{Cosmology and Astrophysics}
 
 The talks by Reddy and Bojowald addressed fundamental and deep problems of great importance to physics and to our field of research..
 \begin{itemize}
 \item{  How do we properly formulate a theory of gravity?}
 \item{What are the phase transitions in cosmology and how to they determine the world in which we live?}
 \item{What is the nature of dark energy and dark matter?}
 \item{How was the baryon asymmetry generated?}
 \end{itemize}
 The solutions to these problems no doubt go outside of the theory of strong interactions.  They are nevertheless very important and people in our field should think about them.
 
 There are a set of problems in high energy astrophysics well within our range of study:  
 \begin{itemize}
 \item{Neutron stars, and their phenomenology}
 \item{Gamma ray bursts and the formation of lack holes}
 \item{Active galactic nuclei}
 \end{itemize}
 
 Reddy showed that recent measurements of neutron star masses are beginning to put limits on the equation of state for cold matter at very high energy density.  These limits may ultimately
 challenge quark matter models of neutron star interiors.

\section*{Acknowledgments}

The research of  L. McLerran is supported under DOE Contract No. DE-AC02-98CH10886.


\begin{thebibliography}{00}

\bibitem{Stasto:2000er}
  A.~M.~Stasto, K.~J.~Golec-Biernat and J.~Kwiecinski,
  Phys.\ Rev.\ Lett.\  {\bf 86}, 596 (2001)
  [arXiv:hep-ph/0007192].
  
\bibitem{Khachatryan:2010xs}
  V.~Khachatryan {\it et al.}  [CMS Collaboration],
  JHEP {\bf 1002}, 041 (2010)
  [arXiv:1002.0621 [hep-ex]].
  
\bibitem{Khachatryan:2010us}
  V.~Khachatryan {\it et al.}  [CMS Collaboration],
  Phys.\ Rev.\ Lett.\  {\bf 105}, 022002 (2010)
  [arXiv:1005.3299 [hep-ex]].

\bibitem{McLerran:2010ex}
  L.~McLerran and M.~Praszalowicz,
  Acta Phys.\ Polon.\  B {\bf 41}, 1917 (2010)
  [arXiv:1006.4293 [hep-ph]].

\bibitem{McLerran:2010wm}
  L.~McLerran and M.~Praszalowicz,
  Acta Phys.\ Polon.\  B {\bf 42}, 99 (2011)
  [arXiv:1011.3403 [hep-ph]].
  
\bibitem{Tribedy:2011yn}
  P.~Tribedy and R.~Venugopalan,
  arXiv:1101.5922 [hep-ph].

\bibitem{Abe:1988yu}
  F.~Abe {\it et al.}  [CDF Collaboration],
   ``Transverse momentum distributions of charged particles produced in
  Phys.\ Rev.\ Lett.\  {\bf 61}, 1819 (1988).

\bibitem{Aamodt:2010pp}
  K.~Aamodt {\it et al.}  [ALICE Collaboration],
   ``Charged-particle multiplicity measurement in proton-proton collisions at
  Eur.\ Phys.\ J.\  C {\bf 68}, 345 (2010)
  [arXiv:1004.3514 [hep-ex]].

\bibitem{Aad:2010rd}
  G.~Aad {\it et al.}  [ATLAS Collaboration],
   ``Charged-particle multiplicities in pp interactions at sqrt(s) = 900 GeV
  Phys.\ Lett.\  B {\bf 688}, 21 (2010)
  [arXiv:1003.3124 [hep-ex]].

\bibitem{Braidot:2010zh}
  E.~Braidot  [STAR Collaboration],
  arXiv:1005.2378 [hep-ph].
  
\bibitem{Albacete:2010pg}
  J.~L.~Albacete and C.~Marquet,
  Phys.\ Rev.\ Lett.\  {\bf 105}, 162301 (2010)
  [arXiv:1005.4065 [hep-ph]].
  
\bibitem{Tuchin:2009nf}
  K.~Tuchin,
  Nucl.\ Phys.\  A {\bf 846}, 83 (2010)
  [arXiv:0912.5479 [hep-ph]].
  
\bibitem{Aamodt:2010pb}
  K.~Aamodt {\it et al.}  [The ALICE Collaboration],
  Phys.\ Rev.\ Lett.\  {\bf 105}, 252301 (2010)
  [arXiv:1011.3916 [nucl-ex]].
  
\bibitem{Dusling:2010rm}
  K.~Dusling, T.~Epelbaum, F.~Gelis and R.~Venugopalan,
  Nucl.\ Phys.\  A {\bf 850}, 69 (2011)
  [arXiv:1009.4363 [hep-ph]].
  
\bibitem{Policastro:2001yc}
  G.~Policastro, D.~T.~Son and A.~O.~Starinets,
  Phys.\ Rev.\ Lett.\  {\bf 87}, 081601 (2001)
  [arXiv:hep-th/0104066].

\bibitem{Asakawa:2010xf}
  M.~Asakawa, S.~A.~Bass and B.~Muller,
  arXiv:1008.3496 [nucl-th].
  
\bibitem{:2009dqa}
  B.~I.~Abelev {\it et al.}  [STAR Collaboration],
   ``Growth of Long Range Forward-Backward Multiplicity Correlations with
  Phys.\ Rev.\ Lett.\  {\bf 103}, 172301 (2009)
  [arXiv:0905.0237 [nucl-ex]].
  
\bibitem{Lappi:2009vb}
  T.~Lappi and L.~McLerran,
  Nucl.\ Phys.\  A {\bf 832}, 330 (2010)
  [arXiv:0909.0428 [hep-ph]].

\bibitem{Adams:2004pa}
  J.~Adams {\it et al.}  [STAR Collaboration],
   ``Minijet deformation and charge-independent angular correlations on momentum
  Phys.\ Rev.\  C {\bf 73}, 064907 (2006)
  [arXiv:nucl-ex/0411003].

\cite{:2008cqb}
\bibitem{:2008cqb}
  A.~Adare {\it et al.}  [PHENIX Collaboration],
   ``Dihadron azimuthal correlations in Au+Au collisions at $\sqrt(s_NN)=200$
  Phys.\ Rev.\  C {\bf 78}, 014901 (2008)
  [arXiv:0801.4545 [nucl-ex]].

\bibitem{:2009qa}
  B.~I.~Abelev {\it et al.}  [STAR Collaboration],
   ``Long range rapidity correlations and jet production in high energy nuclear
  Phys.\ Rev.\  C {\bf 80}, 064912 (2009)
  [arXiv:0909.0191 [nucl-ex]].
  
\bibitem{Alver:2009id}
  B.~Alver {\it et al.}  [PHOBOS Collaboration],
   ``High transverse momentum triggered correlations over a large pseudorapidity
  Phys.\ Rev.\ Lett.\  {\bf 104}, 062301 (2010)
  [arXiv:0903.2811 [nucl-ex]].

\bibitem{Dumitru:2008wn}
  A.~Dumitru, F.~Gelis, L.~McLerran and R.~Venugopalan,
  Nucl.\ Phys.\  A {\bf 810}, 91 (2008)
  [arXiv:0804.3858 [hep-ph]].


\bibitem{Khachatryan:2010gv}
  V.~Khachatryan {\it et al.}  [CMS Collaboration],
   ``Observation of Long-Range Near-Side Angular Correlations in Proton-Proton
  JHEP {\bf 1009}, 091 (2010)
  [arXiv:1009.4122 [hep-ex]].


\bibitem{Alver:2010gr}
  B.~Alver and G.~Roland,
  Phys.\ Rev.\  C {\bf 81}, 054905 (2010)
  [Erratum-ibid.\  C {\bf 82}, 039903 (2010)]
  [arXiv:1003.0194 [nucl-th]].
  
\bibitem{Gelis:2009wh}
  F.~Gelis, T.~Lappi and L.~McLerran,
  Nucl.\ Phys.\  A {\bf 828}, 149 (2009)
  [arXiv:0905.3234 [hep-ph]].
  
\bibitem{McLerran:2007qj}
  L.~McLerran and R.~D.~Pisarski,
  Nucl.\ Phys.\  A {\bf 796}, 83 (2007)
  [arXiv:0706.2191 [hep-ph]].
  
\bibitem{Kojo:2009ha}
  T.~Kojo, Y.~Hidaka, L.~McLerran and R.~D.~Pisarski,
  Nucl.\ Phys.\  A {\bf 843}, 37 (2010)
  [arXiv:0912.3800 [hep-ph]].

\bibitem{Andronic:2009gj}
  A.~Andronic {\it et al.},
  Nucl.\ Phys.\  A {\bf 837}, 65 (2010)
  [arXiv:0911.4806 [hep-ph]].

\bibitem{Rajagopal:1992qz}
  K.~Rajagopal and F.~Wilczek,
  Nucl.\ Phys.\  B {\bf 399}, 395 (1993)
  [arXiv:hep-ph/9210253].

\bibitem{Huovinen:2009yb}
  P.~Huovinen and P.~Petreczky,
  Nucl.\ Phys.\  A {\bf 837}, 26 (2010)
  [arXiv:0912.2541 [hep-ph]].
  
\bibitem{CasalderreySolana:2010eh}
  J.~Casalderrey-Solana, J.~G.~Milhano and U.~A.~Wiedemann,
  J.\ Phys.\ G {\bf 38}, 035006 (2011)
  [arXiv:1012.0745 [hep-ph]].

\bibitem{Adare:2009qk}
  A.~Adare {\it et al.}  [PHENIX Collaboration],
  Phys.\ Rev.\  C {\bf 81}, 034911 (2010)
  [arXiv:0912.0244 [nucl-ex]].


\bibitem{:2008fqa}
  A.~Adare {\it et al.}  [PHENIX Collaboration],
  Phys.\ Rev.\ Lett.\  {\bf 104}, 132301 (2010)
  [arXiv:0804.4168 [nucl-ex]].







\end{thebibliography}
\end{document}